\input harvmac
\input graphicx

\def\Title#1#2{\rightline{#1}\ifx\answ\bigans\nopagenumbers\pageno0\vskip1in
\else\pageno1\vskip.8in\fi \centerline{\titlefont #2}\vskip .5in}
%

%%%%%%%%%%%%%%%%%%
%
% Figure macros, SBG 3/03
%
\ifx\includegraphics\UnDeFiNeD\message{(NO graphicx.tex, FIGURES WILL BE IGNORED)}
\def\figin#1{\vskip2in}% blank space instead
\else\message{(FIGURES WILL BE INCLUDED)}\def\figin#1{#1}
\fi
\def\Fig#1{Fig.~\the\figno\xdef#1{Fig.~\the\figno}\global\advance\figno
 by1}
%
%  Ifig   usage:
%
%         \Ifig{\Fig\figlabel}{caption}{figfile}{hsize}
%
% where vsize is the desired vertical size of the figure in truein
%
\def\Ifig#1#2#3#4{
\goodbreak\midinsert
\figin{\centerline{
\includegraphics[width=#4truein]{#3}}}
\narrower\narrower\noindent{\footnotefont
{\bf #1:}  #2\par}
\endinsert
}
%
%defs
%
\font\ticp=cmcsc10
\def\subsubsec#1{\noindent{\undertext {#1}}}
\def\undertext#1{$\underline{\smash{\hbox{#1}}}$}

\def\q{{\bf q}}
\def\x{{\bf x}}

\def\roughly#1{\mathrel{\raise.3ex\hbox{$#1$\kern-.75em\lower1ex\hbox{$\sim$}}}}

\def\Ei{{\rm Ei}}
%
%refs
%

\lref\GiPo{
  S.~B.~Giddings and R.~A.~Porto,
  ``The gravitational S-matrix,''
  Phys.\ Rev.\  D {\bf 81}, 025002 (2010)
  [arXiv:0908.0004 [hep-th]].
  %%CITATION = PHRVA,D81,025002;%%
}
\lref\BaFi{
  T.~Banks and W.~Fischler,
  ``A model for high energy scattering in quantum gravity,''
  arXiv:hep-th/9906038.
  %%CITATION = HEP-TH/9906038;%%
}
\lref\GiTh{
  S.~B.~Giddings and S.~D.~Thomas,
  ``High energy colliders as black hole factories: The end of short  distance
  physics,''
  Phys.\ Rev.\  D {\bf 65}, 056010 (2002)
  [arXiv:hep-ph/0106219].
  %%CITATION = PHRVA,D65,056010;%%
}
\lref\EaGi{
  D.~M.~Eardley and S.~B.~Giddings,
  ``Classical black hole production in high-energy collisions,''
  Phys.\ Rev.\  D {\bf 66}, 044011 (2002)
  [arXiv:gr-qc/0201034].
  %%CITATION = PHRVA,D66,044011;%%
}
\lref\BDF{
  Z.~Bern, L.~J.~Dixon and R.~Roiban,
  ``Is N = 8 Supergravity Ultraviolet Finite?,''
  Phys.\ Lett.\  B {\bf 644}, 265 (2007)
  [arXiv:hep-th/0611086].
  %%CITATION = PHLTA,B644,265;%%
}
\lref\DixonGZ{
  L.~J.~Dixon,
  ``Ultraviolet Behavior of N=8 Supergravity,''
  arXiv:1005.2703 [hep-th].
  %%CITATION = ARXIV:1005.2703;%%
}
\lref\GoRo{
  W.~D.~Goldberger and I.~Z.~Rothstein,
  ``An effective field theory of gravity for extended objects,''
  Phys.\ Rev.\  D {\bf 73}, 104029 (2006)
  [arXiv:hep-th/0409156].
  %%CITATION = PHRVA,D73,104029;%%
}
\lref\ACVone{
  D.~Amati, M.~Ciafaloni and G.~Veneziano,
  ``Superstring Collisions at Planckian Energies,''
  Phys.\ Lett.\  B {\bf 197}, 81 (1987)\semi 
  %%CITATION = PHLTA,B197,81;%%
  D.~Amati, M.~Ciafaloni and G.~Veneziano,
  ``Classical and Quantum Gravity Effects from Planckian Energy Superstring
  Collisions,''
  Int.\ J.\ Mod.\ Phys.\  A {\bf 3}, 1615 (1988).
  %%CITATION = IMPAE,A3,1615;%%
}
\lref\ACVtwo{
  D.~Amati, M.~Ciafaloni and G.~Veneziano,
  ``Can Space-Time Be Probed Below The String Size?,''
  Phys.\ Lett.\  B {\bf 216}, 41 (1989).
  %%CITATION = PHLTA,B216,41;%%
}
\lref\ACVthree{
  D.~Amati, M.~Ciafaloni and G.~Veneziano,
  ``Higher Order Gravitational Deflection And Soft Bremsstrahlung In Planckian
  Energy Superstring Collisions,''
  Nucl.\ Phys.\  B {\bf 347}, 550 (1990).
  %%CITATION = NUPHA,B347,550;%%
}
\lref\ACVfour{
  D.~Amati, M.~Ciafaloni and G.~Veneziano,
  ``Effective action and all order gravitational eikonal at Planckian
  energies,''
  Nucl.\ Phys.\  B {\bf 403}, 707 (1993).
  %%CITATION = NUPHA,B403,707;%%
}
\lref\ASrev{
  M.~Niedermaier and M.~Reuter,
  ``The Asymptotic Safety Scenario in Quantum Gravity,''
  Living Rev.\ Rel.\  {\bf 9}, 5 (2006).
  %%CITATION = 00222,9,5;%%
}
\lref\AiSe{
  P.~C.~Aichelburg and R.~U.~Sexl,
  ``On the Gravitational field of a massless particle,''
  Gen.\ Rel.\ Grav.\  {\bf 2}, 303 (1971).
  %%CITATION = GRGVA,2,303;%%
}
\lref\WeinAS{S.~Weinberg, ``Ultraviolet Divergences In Quantum Theories Of Gravitation," in {\sl General Relativity: An Einstein centenary survey,} eds. S. W. Hawking and W. Israel, Cambridge University Press (1979).}
\lref\Astrorev{
  A.~Strominger,
  ``Les Houches lectures on black holes,''
  arXiv:hep-th/9501071.
  %%CITATION = HEP-TH 9501071;%%
}
\lref\SGinfo{S.~B.~Giddings,
  ``Quantum mechanics of black holes,''
  arXiv:hep-th/9412138\semi
  %%CITATION = HEP-TH 9412138;%%
  ``The black hole information paradox,''
  arXiv:hep-th/9508151.
  %%CITATION = HEP-TH 9508151;%%
}
\lref\GMH{
  S.~B.~Giddings, D.~Marolf and J.~B.~Hartle,
  ``Observables in effective gravity,''
  Phys.\ Rev.\  D {\bf 74}, 064018 (2006)
  [arXiv:hep-th/0512200].
  %%CITATION = PHRVA,D74,064018;%%
}
\lref\GaGi{
  M.~Gary and S.~B.~Giddings,
  ``Relational observables in 2d quantum gravity,''
  arXiv:hep-th/0612191, Phys.\ Rev.\  D {\bf 75} 104007 (2007).
  %%CITATION = HEP-TH/0612191;%%
}
\lref\GiSr{
  S.~B.~Giddings and M.~Srednicki,
  ``High-energy gravitational scattering and black hole resonances,''
  Phys.\ Rev.\  D {\bf 77}, 085025 (2008)
  [arXiv:0711.5012 [hep-th]].
  %%CITATION = PHRVA,D77,085025;%%
}
\lref\KabatTB{
  D.~N.~Kabat and M.~Ortiz,
  ``Eikonal Quantum Gravity And Planckian Scattering,''
  Nucl.\ Phys.\  B {\bf 388}, 570 (1992)
  [arXiv:hep-th/9203082].
  %%CITATION = NUPHA,B388,570;%%
}
\lref\DittrichVV{
  W.~Dittrich,
  ``Equivalence of the dirac equation to a subclass of feynman diagrams,''
  Phys.\ Rev.\  D {\bf 1}, 3345 (1970).
  %%CITATION = PHRVA,D1,3345;%%
}
\lref\DittrichBE{
  W.~Dittrich,
  ``Bloch-Nordsieck approximation in linearized quantum gravity,''
  arXiv: hep-th/0010235.
  %%CITATION = HEP-TH/0010235;%%
}
\lref\VeVe{
  H.~L.~Verlinde and E.~P.~Verlinde,
  ``Scattering at Planckian Energies,''
  Nucl.\ Phys.\  B {\bf 371}, 246 (1992)
  [arXiv:hep-th/9110017].
  %%CITATION = NUPHA,B371,246;%%
}
\lref\Banks{
  T.~Banks,
  ``A critique of pure string theory: Heterodox opinions of diverse
  dimensions,''
  arXiv:hep-th/0306074.
  %%CITATION = HEP-TH/0306074;%%
}
\lref\BernDixetal{
  Z.~Bern, L.~J.~Dixon, D.~C.~Dunbar, M.~Perelstein and J.~S.~Rozowsky,
  ``On the relationship between Yang-Mills theory and gravity and its
  implication for ultraviolet divergences,''
  Nucl.\ Phys.\  B {\bf 530}, 401 (1998)
  [arXiv:hep-th/9802162].
  %%CITATION = NUPHA,B530,401;%%
}
\lref\Page{
  D.~N.~Page,
  ``Information in black hole radiation,''
  Phys.\ Rev.\ Lett.\  {\bf 71}, 3743 (1993)
  [arXiv:hep-th/9306083].
  %%CITATION = PRLTA,71,3743;%%
}
\lref\QBHB{
  S.~B.~Giddings,
  ``Quantization in black hole backgrounds,''
  Phys.\ Rev.\  D {\bf 76}, 064027 (2007)
  [arXiv:hep-th/0703116].
  %%CITATION = PHRVA,D76,064027;%%
}
\lref\Hawkunc{
  S.~W.~Hawking,
  ``Breakdown Of Predictability In Gravitational Collapse,''
  Phys.\ Rev.\  D {\bf 14}, 2460 (1976).
  %%CITATION = PHRVA,D14,2460;%%
}
\lref\LQGST{
  S.~B.~Giddings,
  ``Locality in quantum gravity and string theory,''
  Phys.\ Rev.\  D {\bf 74}, 106006 (2006)
  [arXiv:hep-th/0604072].
  %%CITATION = PHRVA,D74,106006;%%
}
\lref\GGM{
  S.~B.~Giddings, D.~J.~Gross and A.~Maharana,
  ``Gravitational effects in ultrahigh-energy string scattering,''
  Phys.\ Rev.\  D {\bf 77}, 046001 (2008)
  [arXiv:0705.1816 [hep-th]].
  %%CITATION = PHRVA,D77,046001;%%
}
\lref\GiddingsSJ{
  S.~B.~Giddings,
  ``Black hole information, unitarity, and nonlocality,''
  Phys.\ Rev.\  D {\bf 74}, 106005 (2006)
  [arXiv:hep-th/0605196].
  %%CITATION = PHRVA,D74,106005;%%
}
\lref\GiddingsBE{
  S.~B.~Giddings,
  ``(Non)perturbative gravity, nonlocality, and nice slices,''
  Phys.\ Rev.\  D {\bf 74}, 106009 (2006)
  [arXiv:hep-th/0606146].
  %%CITATION = PHRVA,D74,106009;%%
}
\lref\Weinsoft{
  S.~Weinberg,
  ``Infrared photons and gravitons,''
  Phys.\ Rev.\  {\bf 140}, B516 (1965).
  %%CITATION = PHRVA,140,B516;%%
}
\lref\DiLa{
  S.~Dimopoulos and G.~L.~Landsberg,
  ``Black Holes at the LHC,''
  Phys.\ Rev.\ Lett.\  {\bf 87}, 161602 (2001)
  [arXiv:hep-ph/0106295].
  %%CITATION = PRLTA,87,161602;%%
}
\lref\SGPascos{
  S.~B.~Giddings,
  ``High-energy black hole production,''
  AIP Conf.\ Proc.\  {\bf 957}, 69 (2007)
  [arXiv:0709.1107 [hep-ph]].
  %%CITATION = APCPC,957,69;%%
}
\lref\Duff{ M.~J.~Duff,
  ``Covariant gauges and point sources in general relativity,''
  Annals Phys.\  {\bf 79}, 261 (1973)\semi
  %%CITATION = APNYA,79,261;%% 
  ``Quantum tree graphs and the Schwarzschild solution," Phys.\ Rev.\ {\bf D7} 2317 (1973).}
\lref\Rych{
  P.~Lodone and S.~Rychkov,
  ``Radiation Problem in Transplanckian Scattering,''
  JHEP {\bf 0912}, 036 (2009)
  [arXiv:0909.3519 [hep-ph]].
  %%CITATION = JHEPA,0912,036;%%
}
\lref\Empar{
  R.~Emparan, M.~Masip and R.~Rattazzi,
  ``Cosmic rays as probes of large extra dimensions and TeV gravity,''
  Phys.\ Rev.\  D {\bf 65}, 064023 (2002)
  [arXiv:hep-ph/0109287].
  %%CITATION = PHRVA,D65,064023;%%
}
\lref\Venez{
  G.~Veneziano,
  ``String-theoretic unitary S-matrix at the threshold of black-hole
  production,''
  JHEP {\bf 0411}, 001 (2004)
  [arXiv:hep-th/0410166].
  %%CITATION = JHEPA,0411,001;%%
}
\lref\BeyP{
  S.~B.~Giddings,
  ``Beyond the Planck scale,''
  arXiv:0910.3140 [gr-qc].
  %%CITATION = ARXIV:0910.3140;%%
}
\lref\GiddingsSJ{
  S.~B.~Giddings,
  ``Black hole information, unitarity, and nonlocality,''
  Phys.\ Rev.\  D {\bf 74}, 106005 (2006)
  [arXiv:hep-th/0605196].
  %%CITATION = PHRVA,D74,106005;%%
}
\lref\GiddingsBE{
  S.~B.~Giddings,
  ``(Non)perturbative gravity, nonlocality, and nice slices,''
  Phys.\ Rev.\  D {\bf 74}, 106009 (2006)
  [arXiv:hep-th/0606146].
  %%CITATION = PHRVA,D74,106009;%%
}
\lref\Penrose{R. Penrose {\sl unpublished} (1974).}
\lref\GSP{S.B. Giddings, R. Porto, and M. Schmidt-Sommerfeld, under investigation.}
\lref\GiddingsPT{
  S.~B.~Giddings and M.~Lippert,
  ``Precursors, black holes, and a locality bound,''
  Phys.\ Rev.\  D {\bf 65}, 024006 (2002)
  [arXiv:hep-th/0103231]\semi
  ``The information paradox and the locality bound,''
  Phys.\ Rev.\  D {\bf 69}, 124019 (2004)
  [arXiv:hep-th/0402073].
  %%CITATION = PHRVA,D69,124019;%%
  %%CITATION = PHRVA,D65,024006;%%
}
\lref\GiSl{
  S.~B.~Giddings and M.~S.~Sloth,
  ``Semiclassical relations and IR effects in de Sitter and slow-roll
  space-times,''
  arXiv:1005.1056 [hep-th].
  %%CITATION = ARXIV:1005.1056;%%
}
\lref\SGuvsc{
  S.~B.~Giddings,
  ``Nonlocality vs. complementarity: a conservative approach to the information
  problem,''
  arXiv:0911.3395 [hep-th].
  %%CITATION = ARXIV:0911.3395;%%
}
\lref\GiRy{
  S.~B.~Giddings and V.~S.~Rychkov,
  ``Black holes from colliding wavepackets,''
  Phys.\ Rev.\  D {\bf 70}, 104026 (2004)
  [arXiv:hep-th/0409131].
  %%CITATION = PHRVA,D70,104026;%%
}
\lref\MuzinichIN{
  I.~J.~Muzinich and M.~Soldate,
  ``High-Energy Unitarity of Gravitation and Strings,''
  Phys.\ Rev.\  D {\bf 37}, 359 (1988).
  %%CITATION = PHRVA,D37,359;%%
}
\lref\Schnone{
  H.~J.~Schnitzer,
 ``Reggeization of N=8 Supergravity and N=4 Yang-Mills Theory II,''
  arXiv:0706.0917 [hep-th].
  %%CITATION = ARXIV:0706.0917;%%
}
\lref\Schntwo{
  H.~J.~Schnitzer,
  ``Reggeization of N = 8 supergravity and N = 4 Yang-Mills theory,''
  arXiv:hep-th/0701217.
  %%CITATION = HEP-TH/0701217;%%
}

\Title{
\vbox{
%\hbox{Draft -- do not distribute}
\hbox{CERN-PH-TH/2010-122}
\hbox{NSF-KITP-10-074}}
\vbox{\baselineskip12pt
}}
{\vbox{\centerline{High energy scattering in gravity and supergravity}
}}
\vskip-.05in
\centerline{{\ticp Steven B. Giddings,${}^a$\footnote{$^\ast$}{Email address: giddings@physics.ucsb.edu}  Maximilian Schmidt-Sommerfeld,${}^{b,c}$\footnote{$^\dagger$}{Email address: maximilian.schmidt-sommerfeld@cern.ch} and  Jeppe R. Andersen$^b$\footnote{$^\ddagger$}{Email address: jeppe.andersen@cern.ch  }} }
\vskip.1in
\centerline{\sl ${}^a$Department of Physics, University of California, Santa Barbara, CA 93106}
\vskip.1in
\centerline{\sl ${}^b$PH-TH, CERN, 1211 Geneve 23, Switzerland}
\vskip.1in
\centerline{\sl ${}^c$Kavli Institute for Theoretical Physics, University of California, Santa Barbara, CA 93106}
%
%\centerline{{\ticp Steven B. Giddings,\footnote{$^\ast$}{Email address: giddings@physics.ucsb.edu}  } }
%\centerline{\sl Department of Physics, University of California, Santa Barbara, CA 93106}
%\vskip.1in
%\centerline{{\ticp Maximilian Schmidt-Sommerfeld,\footnote{$^\dagger$}{Email address: maximilian.schmidt-sommerfeld@cern.ch} }}
%\centerline{\sl PH-TH, CERN, 1211 Geneve 23, Switzerland}
%\centerline{\sl Kavli Institute for Theoretical Physics, Kohn Hall, UCSB, Santa Barbara, CA 93106}
%\vskip.1in
%\centerline{{\ticp and Jeppe R. Andersen\footnote{$^\ddagger$}{Email address: jeppe.andersen@cern.ch  } }}
%\centerline{\sl PH-TH, CERN, 1211 Geneve 23, Switzerland}
\vskip.15in
\centerline{\bf Abstract}
We investigate features of perturbative gravity and supergravity by studying scattering in the ultraplanckian limit, and sharpen arguments that the dynamics is governed by long-distance physics.  A simple example capturing aspects of the eikonal resummation suggests why short distance phenomena and in particular divergences or nonrenormalizability do not necessarily play a central role in this regime.  A more profound problem is apparently unitarity.  These considerations can be illustrated by showing that known gravity and supergravity amplitudes have the same long-distance behavior, despite the extra light states of supergravity, and this serves as an important check on long-range dynamics in a context where perturbative amplitudes are finite.  We also argue that these considerations have other important implications:  they obstruct probing the conjectured phenomenon of asymptotic safety through a physical scattering process, and  ultraplanckian scattering exhibiting Regge behavior. These arguments sharpen the need to find a nonperturbative completion of gravity with mechanisms which restore unitarity in the strong gravity regime.

%\draftmode
\Date{}

\newsec{Introduction}

One of the challenges in finding a quantum theory of gravity is to formulate sharp physical questions.  In particular, we cannot define the theory through gauge invariant correlators of local operators, unlike with other field theories.  In formulating well-defined quantities, one apparently has two choices.  One is to define the gravitational S-matrix, or a modification of it accounting for soft graviton divergences, analogous to those in QED. Another is to formulate appropriate observables that are approximately local, under certain conditions on the state, {\it etc.}; we expect such observables to be relevant in cases where a description is needed that doesn't directly refer to asymptotic infinity, as in studies of quantum cosmology.

Nonrenormalizability has been regarded as another central problem of quantum gravity, and in particular has been a strong motivator for the development of string theory, as well as for the study of supergravity (SUGRA) and its order-by-order finiteness\refs{\DixonGZ}.  While some progress has been made on formulating approximately local observables in relational approaches\refs{\GMH,\GaGi}, we will focus on the relevance of (non)renormalizability to the problem of finding the S-matrix.  

Some properties of the S-matrix, such as its analyticity structure, crossing, and locality,
were investigated accounting for  essential aspects of gravity in \GiPo.  A particularly interesting regime -- that must be described by any theory of gravity with asymptotically flat solutions and Lorentz invariance -- is that of scattering at energies far beyond the Planck scale.  This is because Lorentz invariance, together with a very mild notion of locality, allows us to consider asymptotic states with arbitrarily high energies, that may then undergo an ultraplanckian collision.

It is natural to investigate the role of renormalizability in this regime.  In particular, if individual loop amplitudes are rendered finite by string theory or supergravity, is this all that is needed to ensure that the high-energy S-matrix can be derived?  Or, following another proposal, might a phenomenon such as asymptotic safety\refs{\WeinAS,\ASrev} be critical in formulating quantum gravity?

A counterargument to this is that ultrahigh-energy scattering in gravity in fact probes long-distance physics; this phenomenon has been investigated in various works, particularly \refs{\BaFi\GiTh\EaGi\Banks-\LQGST,\GiPo}.  Masslessness of the graviton together with growth of the dimensionless gravitational coupling  with energy are responsible for this behavior.  If indeed ultra-high energy gravitational scattering is governed by long distance dynamics, this suggests that short-distance singularities in loop amplitudes are not centrally relevant.  Moreover, at ultrahigh-energies, and sufficiently low impact parameter, classical gravity predicts black hole formation\refs{\Penrose,\EaGi}.  And, perturbative quantization about such a solution has been argued by Hawking\Hawkunc\ to violate unitary evolution.  Combining these facts suggests that in gravity, unitarity is a more profound question than renormalizability or finiteness\refs{\BeyP}.\foot{In electroweak theory, these two questions are closely connected.  However, the feature that in gravity high energies probe long distances, as we will explore further, appears to significantly weaken such a link.}

A first reaction to this story might be that we cannot sharply discuss the ultrahigh energy regime in any case, because there gravity is strongly coupled.  However, this appears not to be true.  One can precisely describe high energy scattering in the regime of large impact parameters or small momentum transfers -- like in the ultraplanckian scattering of the Earth and the Moon!  Of course, it is conceivable that the relativistic case could introduce new features, so it is important to study its specifics.  But, there, as we will elaborate, at sufficiently large impact parameters amplitudes are well approximated by the Born approximation.  There are also general arguments that for smaller impact parameters, there is still a consistent loop expansion for gravity, producing the amplitudes of the eikonal approximation\refs{\ACVone\ACVtwo\MuzinichIN\VeVe\DittrichVV\DittrichBE-\KabatTB}.

However, when discussing loop amplitudes, one may be uneasy that the perturbative infinities could be important, or in theories where those infinites are removed, the new states that do so are important.  This question was probed for string theory in \refs{\LQGST,\GGM}, where it was argued that excitation of strings does not significantly change the picture of high-energy, long-range gravitational scattering, and in particular does not interfere with the process classically described as black hole formation.

Another place to investigate these questions is in supergravity, particularly in light of arguments that maximal SUGRA may be perturbatively finite in four dimensions\BDF.  Whether or not it is, superpartners regulate divergences at low loop order.  A natural question is whether these new states somehow alter the picture that ultrahigh-energy scattering is dominated by semiclassical physics at sufficiently large distance.  Precise knowledge of certain SUGRA amplitudes  allows explicit checks of a possible role of new states in modifying the physical picture.

In this paper we investigate SUGRA amplitudes and their match to the eikonal amplitudes, and subleading corrections for large impact parameter.  We indeed find that the essential intuition, that graviton exchange dominates this high-energy regime, holds.  We will better understand the nature of the cutoff that SUGRA supplies, and will argue that this cutoff does not play a central role in the calculation of high-energy amplitudes.  This then supports the statement that perturbative finiteness or renormalizability of amplitudes is not a central focus in ultra-high energy scattering.  However, unitarity is.  The apparent loss of information in black hole formation and evaporation is a serious issue, which has been argued to produce a paradox\refs{\Astrorev,\SGinfo}.\foot{Recent arguments for how a paradox is avoided appear in \refs{\QBHB,\SGuvsc,\GiSl}.}  Assuming gravity does have a unitary S-matrix, a critical question is what physical mechanisms act to produce unitary evolution.  The logic of the information ``paradox" indicates that these need to be nonlocal on scales of the black hole in question, which can be macroscopic.  The need for these mechanisms to unitarize gravity in this nonperturbative regime seems a critical guide to the underlying physics\refs{\LQGST,\GiddingsSJ,\GiddingsBE}.

In outline, the next section of this paper studies a toy integral that nicely illustrates the interplay between the long-distance physics of the eikonal amplitudes, and the short distance physics of the cutoff; in this example, infinities in the perturbation series are a red herring.  Section three  reviews the gravitational eikonal amplitudes, and section four applies the lessons of the toy integral to these amplitudes.  Section five then investigates known explicit supergravity amplitudes, at one and two loops, and shows how they match to the eikonal amplitudes in the relevant regime.  It also gives a general argument for why graviton exchange indeed dominates.  

Section six then discusses aspects of the resulting physical picture.  Outside the black hole regime, scattering is expected to be governed by saddlepoints related to collisions of Aichelburg-Sexl shock waves.  Even if the interactions between these involve large momentum transfers, this is due to exchange of many soft gravitons.  The softness of the individual gravitons has important consequences regarding string excitation, if one works in string theory\refs{\LQGST,\GGM}, for the scale of soft radiation, and for the factorization scale in gravitational scattering of hadrons.  It also raises an important question regarding the possible meaning of an ultraviolet fixed point\refs{\WeinAS,\ASrev}, and suggests that ultrahigh-energy scattering does not exhibit Regge behavior. We also make further comments on the problems of finding a unitary description of scattering in the regime where classically a black hole would form.  Here, it appears that perturbation theory fails, and as noted, new mechanisms are likely needed as part of the nonperturbative completion of gravity that unitarizes the theory, while matching onto a semiclassical picture of black hole formation and evaporation.

\newsec{A toy integral}

As a warmup example of the type of expression we will encounter in high-energy gravitational scattering, let us consider the integral
\eqn\toyint{I=\int_0^1db b^3 e^{ig/b^2}\ ,}
where $g$ is a small constant.
The integrand is singular at $b=0$, but this can be managed by replacing the lower limit using a cutoff at $b=\Lambda^{-1}$.  An obvious approach to evaluating this integral is to Taylor expand the exponential; then each term is trivially integrated, giving
\eqn\powerser{I={1-\Lambda^{-4}\over 4} + ig{(1-\Lambda^{-2})\over 2} - {g^2\over 2} \log\Lambda + {ig^3\over 12}(1-\Lambda^2)+\cdots\ .}
However, we of course find that the resulting expression is very badly behaved as the cutoff is removed, $\Lambda\rightarrow\infty$.  This is not to say that the {\it integral} is badly behaved.  In fact, the regulated integral can be found exactly, in terms of known functions:
\eqn\Iint{I(\Lambda) = {g^2\over 4} \left[\Ei(ig) +{1\over g}\left(i+{1\over g}\right)e^{ig} - \Ei(ig\Lambda^2) -{1\over g\Lambda^2}\left(i+{1\over g\Lambda^2}\right)e^{ig\Lambda^2}\right]\ ,}
where \Ei\ is the exponential integral.  Moreover, this expression has a sensible limit as $\Lambda\rightarrow\infty$; using the asymptotic behavior of the exponential integral as $x\rightarrow\infty$, 
\eqn\eiasy{ \Ei(ix)= i\pi +e^{ix} \left[ -{i\over x} - \left( {1\over x}\right)^2 + {\cal O} \left({1\over x^3}\right) \right]  }
gives 
\eqn\Ilim{I(\Lambda) = {g^2\over 4} \left[\Ei(ig) +{1\over g}\left(i+{1\over g}\right)e^{ig}-i\pi\right] +{\cal O}\left({e^{ig\Lambda^2}\over \Lambda^2}\right)\ .}
Clearly, expanding the exponential is not the right thing
to do in order to evaluate the integral.  There appear to be close parallels in expressions describing gravitational scattering.

\newsec{Gravitational eikonal amplitudes}

Consider gravitational scattering, in $D$ dimensions, in the ultrahigh-energy limit, $E={\sqrt s}\gg M_D$, with $M_D$  the Planck mass.  Since the dimensionless gravitational coupling is usually thought to be $G_D E^{D-2}\sim (E/M_D)^{D-2}$, one might expect this to be a strongly-coupled problem.  However, that depends on the size of the momentum transfer, $t=-q^2$, or impact parameter -- scattering at sufficiently large impact parameter is dominated simply by the Born approximation,
\eqn\born{T_{0}(s,t)=-8\pi G_D  s^2/t. }
 For decreasing impact parameter, higher-loop amplitudes become relevant, and one enters regimes where different phenomena dominate; an overview of these regimes, with further references, is provided in \refs{\GiPo}. (Important earlier references include \refs{\ACVone,\ACVtwo,\ACVthree,\ACVfour,\BaFi}.) In particular, it is argued there and in preceding references that the first loop corrections to become important are the ladder and crossed-ladder diagrams, which can be summed to give the eikonal approximation to the amplitude.

\Ifig{\Fig\gravladd}{A ladder diagram with multiple graviton exchange.}{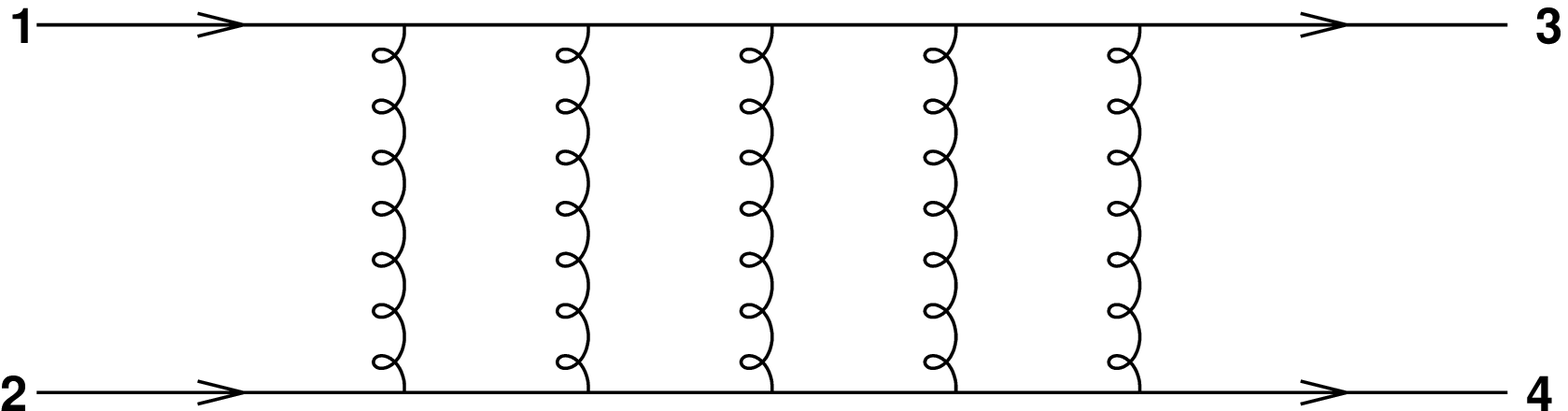}{4}

Specifically, such a ladder diagram is exhibited in \gravladd.  The eikonal approximation to the amplitude arises from neglecting subleading terms in the momentum transfer running through the individual rungs.  In particular, if $k$ denotes a typical such momentum transfer, and $p_i$ an external momentum, then the intermediate propagators of the high-energy particles have denominators of the form
\eqn\denom{D_i = (p_i + k)^2+m^2 = 2p_i\cdot k + k^2\ ,}
and we neglect the second term.  Likewise, in the vertices, we neglect momentum transfers $k\sim q$ compared to the size of $p_i$.  The result is that the sum of ladder and crossed-ladder diagrams at $N$-loop order can be written in terms of the tree amplitude, $T_0(s, -q^2)$ as
\eqn\Nloop{iT_N(s,q) =  {2s\over (N+1)!}\int \left[\prod_{j=1}^{N+1} {d^{D-2}k_j\over (2\pi)^{D-2} }{iT_0(s,-k_j^2)\over 2s} \right] (2\pi)^{D-2} \delta^{D-2}\left(\sum_j k_j - q_\perp\right) \ ,}
where the integrals are over the components of the momenta transverse to those of the incoming particles in the CM frame.  The sum over all such amplitudes gives the eikonal amplitude, which is written in terms of the {\it eikonal phase},
\eqn\eikphase{\eqalign{\chi(x_\perp,s) &= {1\over 2s}\int{d^{D-2}q_\perp\over(2\pi)^{D-2}}\,
e^{i{\bf q}_\perp\cdot\x_\perp}T_{0}(s,-q^2_\perp) \cr
&= {4\pi\over (D-4) \Omega_{D-3}} {G_D s\over x_\perp^{D-4}}\ ,}}
with $\Omega_n$ the area of the unit $n$-sphere.
The eikonal amplitude is
\eqn\Teik{ iT_{\rm eik}(s,t) = 2s \int d^{D-2} x_\perp e^{-i\q_\perp \cdot \x_\perp}(e^{i\chi(x_\perp,s)} -1)\ ;}
in the integral, $b=|x_\perp|$ plays the role of the impact parameter, and thus the amplitude is naturally given in an impact parameter representation.

An important question is to what extent the eikonal amplitude is a good approximation to the exact amplitude, and in what domain.  First, note that a natural expansion parameter is the eikonal phase \eikphase.  When this is small, the eikonal amplitude can be approximated by the linear term in $\chi$, which is exactly the Born amplitude.  Corrections to this become important at impact parameters where $\chi$ becomes of order one.  These impact parameters are directly related to momentum transfers, since the integral \Teik\ has a saddle point which fixes $b$ in terms of $q$.  To write the corresponding equation, we introduce the Schwarzschild radius of the CM energy,
\eqn\schwr{R(E) = {1\over M_D} \left({k_D E\over M_D}\right)^{1/(D-3)},}
where 
\eqn\kddef{k_D = {2(2\pi)^{D-4}\over (D-2) \Omega_{D-2}}.}
The saddle point is then at the the momentum transfer
\eqn\classang{q/E=\theta_c \sim {1 \over E} {\partial \over \partial b} \chi \sim \left[{R(E) \over b}\right]^{D-3}\ ;}
where $\theta_c$ is interpreted as the classical scattering angle.  

\newsec{Corrections to long-distance saddlepoints?}

For momentum transfer given by  \classang\ and such that $\chi>1$, the corrections to Born scattering given by the eikonal amplitude become important.  We wish to know whether and when there are other significant corrections to this amplitude.

There are multiple possible sources of corrections.  First, since  gravity is nonrenormalizable, becoming more so for  $D>4$, there is a question about how to define the $N$-loop amplitudes that we are summing.  There are multiple approaches to making these amplitudes better-behaved in the UV.  One is string theory, which regulates the UV divergences by softening UV behavior through string extendedness.  Another approach is supersymmetry, and it is known that supergravity amplitudes have improved UV behavior, and  even conjectured that in $D=4$ maximal SUGRA these amplitudes could be finite to all loop order\refs{\BernDixetal}.  But, if either of these are true, then there are many more states that can propagate in loops, besides the graviton, and one might be concerned that these make important corrections to the amplitudes \Teik.  Finally, yet another issue is that the approximation of \denom, $D_i\approx 2p_i\cdot k$, worsens the UV behavior of the individual loop amplitudes, since the denominators have decreased order in loop momenta.  

Consider first this last point.  Expanding  \Teik\ in $\chi$, we find that the resulting integrals produce arbitrarily-large inverse powers of a short-distance cutoff on $x_\perp$.  However, the resulting behavior is closely analogous to that of \toyint!   This can in particular be seen by comparing the  case $D=6$.  Given this, one has strong motivation to believe that the divergent behavior in an ultraviolet cutoff $\Lambda$ -- {\it i.e.} at short distances -- is largely a red herring.  Certainly in the case of the example \toyint, one found a finite answer even at infinite cutoff, and indeed the eikonal integral is likewise dominated at the large-distance saddle point given by \classang.  We expect that it is the physics near this saddle point, not at very short distances, that dominates ultra-high energy scattering.

Indeed, this directly addresses also the first concern, since from this viewpoint the infinities contributing to the nonrenormalizability appear not to play a central role.  One sees that the standard gravitational loop expansion about flat space produces expressions that are divergent with the cutoff, but that was also the case with the expansion of the  toy integral, \powerser.  However, we saw this was simply not a useful way to evaluate that integral.

It should be emphasized that the apparent dominance of dynamics near the saddlepoint \classang\ means that in the high-energy regime there is a strong sense in which {\it long distance}, rather than {\it short distance} physics is important.  This is a basic feature of gravity.\foot{Note that this is closely similar to the UV/IR correspondence of string theory, in which high energies produce long strings, and there is some belief in the string community that these are even more directly related through some correspondence between excited string states and black holes.  However, for evidence to the contrary, see \refs{\LQGST,\GGM}, which found a parametric separation between extended string behavior and the long-range behavior of gravity.  For this reason we focus on the gravitational aspects of such behavior.  Earlier references on the black hole aspects of this behavior include \refs{\BaFi\GiTh\EaGi\Banks-\LQGST}, though as we will see in more detail, this behavior also extends outside the black hole regime.}
As noted, there are even deeper issues involving unitarity when one expands about such saddlepoints; we will return to this question later in the paper.

This brings us to the second concern, that states that may be part of the theory, such as excited strings or superpartners could modify the gravitational amplitudes \Teik\ even at long distances.  Possible such modifications due to string behavior were considered in \refs{\LQGST,\GGM}, and while present, have a simple interpretation in terms of tidal excitation of strings, once one reaches small enough impact parameters.  Perhaps more of concern are modifications due to other massless states, such as gravitini, or scalars, or other superpartners that might be exchanged over long distances.

A nice fact is that there has been a lot of study of supergravity amplitudes in the recent literature, and so there are various concrete tests of this possibility.  In particular, since certain higher-loop supergravity amplitudes are known exactly, we can explore whether they are well approximated by the eikonal amplitudes in the regime of ultrahigh energy and small momentum transfer.  This serves as an explicit check on the eikonal amplitudes getting the correct long-distance physics.  We should note that there is a general argument, which we will elaborate later in the paper, that other fields in the gravity multiplet contribute only at subleading order in an expansion in $q/E$:  since it is the highest-helicity field in the multiplet, the graviton couples with the most powers of $E$.  Lower helicity fields have fewer powers of the energy in their couplings, and thus are subdominant at high energies and long distance.  Nonetheless, it is nice to have explicit checks, from exact supergravity amplitudes, that one is not missing important contributions to amplitudes.

It is also interesting to better understand how the supergravity amplitudes regulate the UV divergences order-by-order in the loop expansion, given that 
certain supergravity amplitudes are finite to certain loop orders, and in light of the  all-order conjecture.  One can try to better understand the nature of the supergravity regulation, in comparison with the above discussion, for amplitudes in the high-energy regime.  

But even if supergravity is finite order-by-order, one might ask whether this means that supergravity gives a complete description of the theory.  In light of the preceding discussion, we see that in the very high-energy regime the order-by-order divergences of terms in the loop expansion are not necessarily centrally relevant to the problem of defining the amplitudes.  We will return to this point later in the paper.

\newsec{Supergravity amplitudes and eikonalization}

\subsec{One loop amplitudes}

 \Ifig{\Fig\twograv}{Two graviton exchange.}{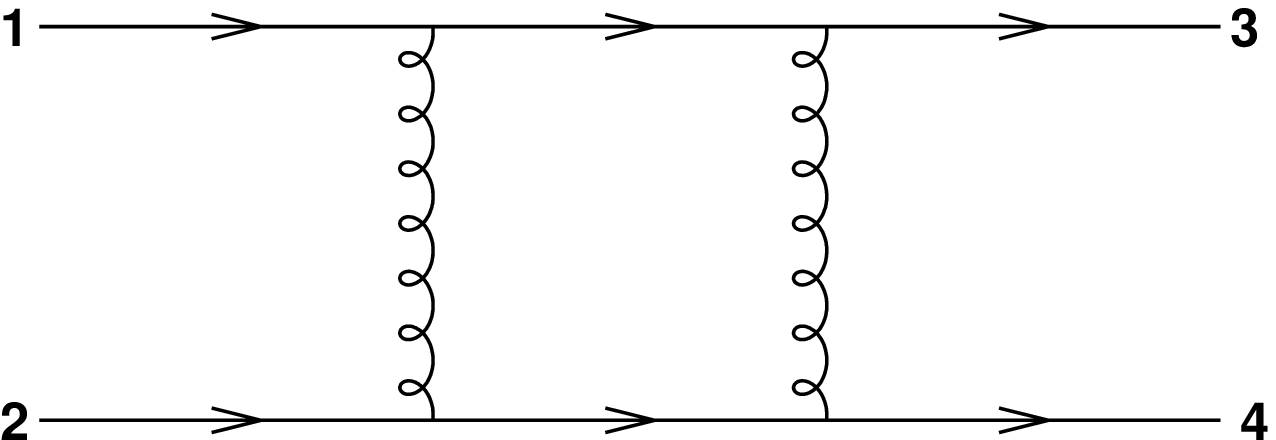}{4}

 \Ifig{\Fig\crossgrav}{A two-graviton crossed ladder.}{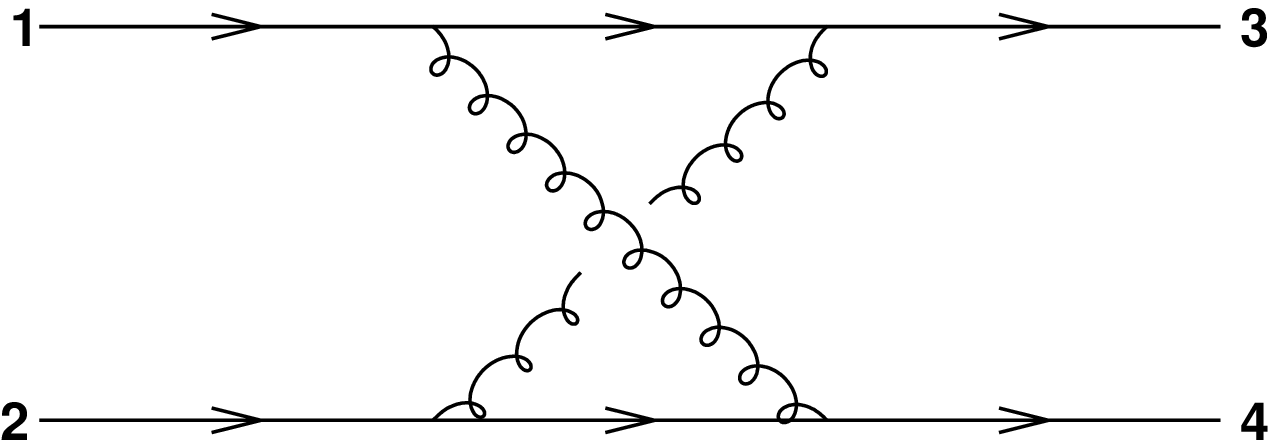}{4}

There has been significant success in explicit calculation of amplitudes in supergravity \refs{\BernDixetal,\DixonGZ}.  Let us consider a four-point amplitude, as pictured in \twograv, corresponding to exchange of two gravitons in the $t$-channel.  In a general gravity theory, not necessarily supersymmetric, the  amplitude will take the form 
\eqn\scalaroneloop{M^{\rm 1-loop}(s,t) = -i\left(8\pi G_D\right)^2 \int {d^D k \over (2\pi)^D}
                   { F(p_i, k) \over k^2 (p_1-k)^2 (p_2+k)^2 (p_1+p_3-k)^2} }
where $F(p_i,k)$ is a polynomial expression that is of order 8 in the external and loop momenta, $p_i$ and $k$ respectively, and can also depend on the polarizations of the particles.  The external states could be scalars, gravitons with definite helicity, or other  states.  To get the full one-loop amplitude, one adds to this the crossed box, pictured in \crossgrav, and other one loop diagrams whose details depend on the structure of the theory.

Focus on the regime $s\gg M_D^2$ and $-t/s\ll1$.  The polynomial $F$ can be expanded in powers of the common high energy, $E=\sqrt s$
\eqn\Fexp{F(p_i,k) = s^4 +\cdots}
where the leading term appears in generic theories. Subsequent terms have fewer powers of $E$, and their details depend on the matter content of the theory.  

In theories where \scalaroneloop\ is divergent for $D<8$, this comes from positive powers of $k$ in the numerator, like for example $k^4$ in 
 $D=4$ gravity coupled to a scalar.   In such a theory, one needs a cutoff in order to regulate the integral.
However, in supersymmetric theories, supersymmetric and other cancellations can remove such divergences.  The {\it exact} one loop gravity amplitude, including {\it all} of the diagrams, takes the form of \scalaroneloop, plus two ``crossed" contributions\refs{\BernDixetal}.  Specifically, defining 
\eqn\oneloopboxintegral{I^{\rm 1}(s,t) = \int {d^D k \over (2\pi)^D}
                        { 1 \over k^2 (p_1-k)^2 (p_2+k)^2 (p_1+p_3-k)^2 }\ , }
the full, convergent (for $D<8$), one-loop N=8 SUGRA amplitude is 
\eqn\sugraoneloop{M_1(s,t) = -i (8 \pi  G_D)^2
                  s^4  
                  \left[ I^{\rm 1}(s,t)+I^{\rm 1}(t,u)+I^{\rm 1}(s,u)\right]\ .}
Thus, any terms of order $k^4$ or higher in the expansion of $F(p_i,k)$ have been eliminated by cancellations.  We can think of the SUGRA theory as providing a regulator.  Since this theory only modifies $F(p_i,k)$ in \scalaroneloop\ significantly when $k\sim \sqrt s$, we see that the regulator is effectively of size $\Lambda\sim \sqrt s$.

\subsec{Match to eikonal}

\Ifig{\Fig\scalarbox}{Scalar box integrals, entering the one-loop supergravity amplitude.}{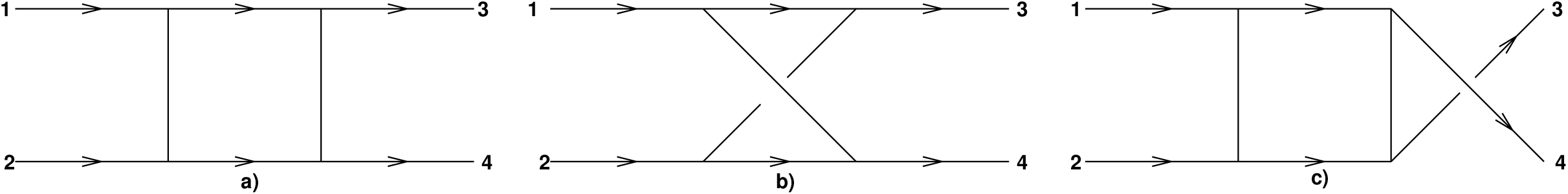}{7}

We can now see how to obtain the eikonal amplitude $T_1$ from the exact supergravity amplitude \sugraoneloop, for $s\gg M_D^2$, $-t/s\ll 1$.  For concreteness, work in the center of mass frame,
\eqn\momentaimcomingparticles{ p_1=(E/2,E/2,{\bf 0})  \qquad p_2=(E/2,-E/2,{\bf 0}) }
and apply a cutoff $\Lambda\roughly<M_D$ to the integral over loop momentum $k$.  In this limit \sugraoneloop\ becomes $-i(8\pi  G_N)^2 s^4$ times the sum of three one-loop integrals corresponding to the three scalar Feynman diagrams given in \scalarbox. In these diagrams, the lines with arrows represent the high energy particles that are being scattered and those without represent the low energy gravitons that the particles exchange.

It is first of all clear that diagram c) is subleading in the circumstances considered, as gravitons with momenta $k\ll M_D$ do not have enough energy-momentum to turn $p_1$ into $p_4=-p_2+q$. Furthermore, given that $k^2\ll M_D^2\ll s$, one can linearize the propagators of the high energy particles as described in section three. With this approximation, the sum of diagrams a) and b) with the appropriate prefactors can be shown 
\refs{\ACVone\ACVtwo\MuzinichIN\VeVe\DittrichVV\DittrichBE-\KabatTB} to be given by $T_1$ in \Nloop. Note that, as expected, the relevant diagrams are the ladder and crossed ladder diagrams.

Thus using an arbitrary large (but not ultralarge) cutoff $\Lambda$ on the loop momentum integrals, the one loop amplitude takes the eikonal form.  We will shortly see that this is true also at two loops.  But first, note that  these can be summed into an expression of the form \Teik, up to corrections small in $t/s$.

One might be concerned that the integrals deviate from their eikonal approximations in the region of large $k$, particularly given that SUGRA supplies the cutoff $\Lambda\sim\sqrt s$.  But, here we return to the discussion of the preceding section.   The sum of the amplitudes gives an integral with similar properties to our toy integral \toyint; thus, as there, we expect that the correct way to evaluate this integral is to expand around the long-distance saddlepoint.  The apparently large contributions as the cutoff is taken to a large value, $\Lambda\rightarrow \sqrt s$ in SUGRA, are apparently direct analogs of the large terms in the expansion \powerser\ -- they, too, seem to be a red herring.  The amplitude in the eikonal regime $\chi\roughly> 1$ should be essentially independent of this short distance cutoff, but this will be apparent only if it is calculated in the correctly summed form.  For large $s$ and small $t$, long distance physics dominates.\foot{Note that in the case $D=4$, the gravitational amplitudes are also IR divergent, but the same divergences occur both in the exact amplitudes and in the eikonal approximation.  They should have the standard remedy, via a sum over soft emitted gravitons.}

\subsec{Two loop amplitudes}

\Ifig{\Fig\twoloop}{Scalar integrals needed for the two-loop supergravity amplitude.}{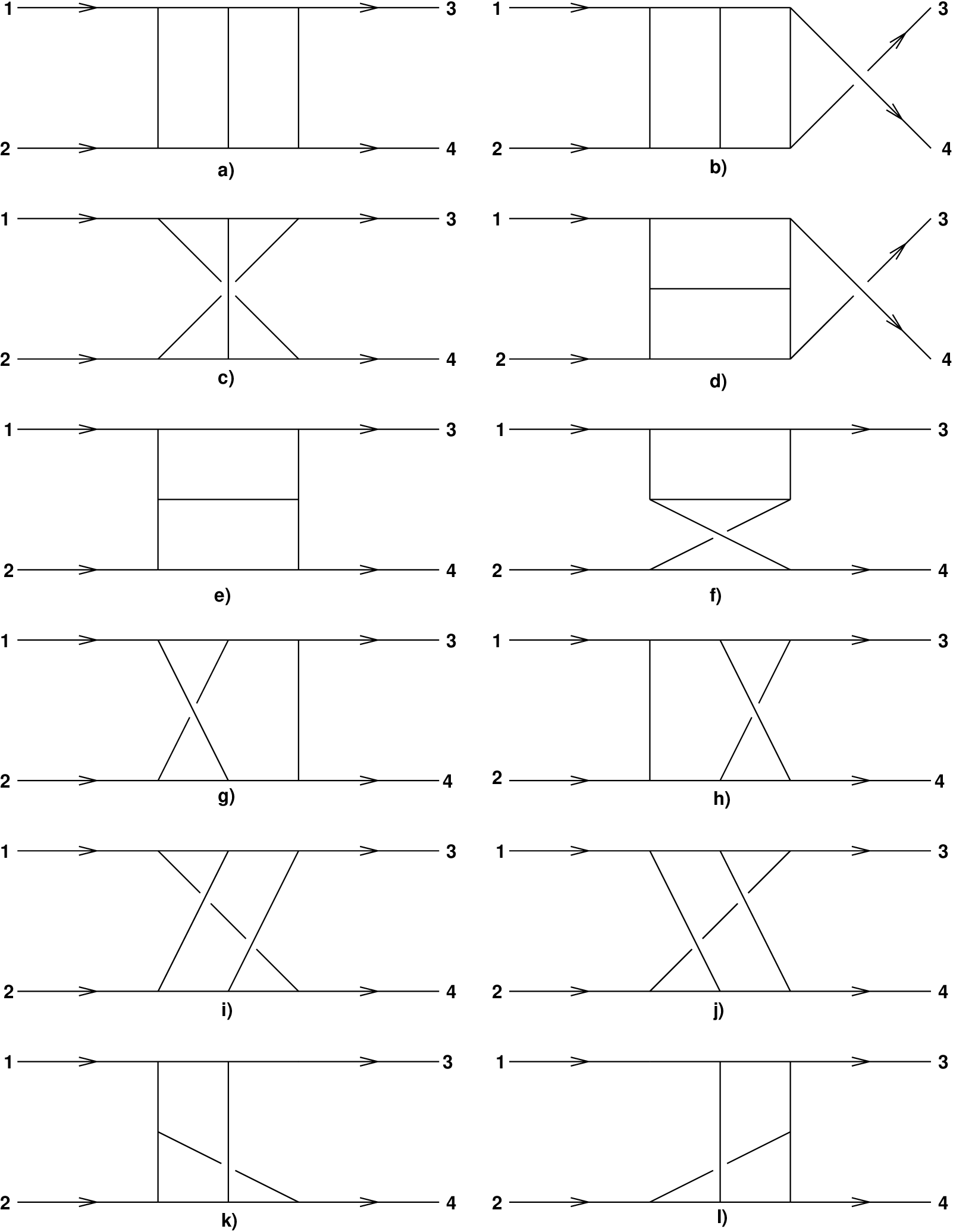}{6}

As a further check, let us consider the two loop case, where things are similar, but more Feynman diagrams need to be considered. The exact maximal supergravity amplitude is, up  to phases and polarization factors depending on the type of the external particles, 
given by \refs{\BernDixetal} 
\eqn\exacttwoloopsugraamplitude{\eqalign{ M_2 = (8\pi G_D)^3 s^4 &\Big[
           							     s^2 I^{2P}(s,t) + s^2 I^{2P}(s,u) +  u^2 I^{2P}(u,t) + u^2 I^{2P}(u,s) 
							               \cr &\phantom{\Big[} +  t^2 I^{2P}(t,s) + t^2 I^{2P}(t,u)
							     	     + s^2 I^{2NP}(s,t) + s^2 I^{2NP}(s,u) 
								     \cr &\phantom{\Big[} +  u^2 I^{2NP}(u,t) + u^2 I^{2NP}(u,s) +  t^2 I^{2NP}(t,s) + t^2 I^{2NP}(t,u) \Big]}}
where 
\eqn\twoloopintegrals{\eqalign{  I^{2P}(s,t) = \int &{d^D k_1 \over (2\pi)^D}{d^D k_2 \over (2\pi)^D} \cr&
						   { 1 \over k_1^2 k_2^2 (k_1+k_2)^2 (k_1-p_1)^2 (k_1-p_1-p_2)^2 (k_2-p_3)^2 (k_2-p_3-p_4)^2} \cr
						     I^{2NP}(s,t) = \int& {d^D k_1 \over (2\pi)^D}{d^D k_2 \over (2\pi)^D} \cr
						   &{ 1 \over k_1^2 k_2^2 (k_1+k_2)^2 (k_1-p_2)^2 (k_1+k_2+p_1)^2 (k_2-p_4)^2 (k_2-p_3-p_4)^2} .} }
A graphical representation of the twelve integrals appearing in \exacttwoloopsugraamplitude\ is given in  \twoloop. In order to obtain the two-loop contribution $T_2$ in \Nloop\ to the eikonal amplitude, one proceeds as at one loop. First of all, given that $-t/s\ll 1$, one can replace the factors of $u^2$ in \exacttwoloopsugraamplitude\ by $s^2$ and drop the terms multiplied by $t^2$. Furthermore, by looking at \twoloop\ one sees that diagrams b) and d), i.e. the second and fourth term in the bracket of \exacttwoloopsugraamplitude, can be neglected as, in analogy to what was said for the third term in \sugraoneloop, the exchanged gravitons do not have enough energy-momentum to turn $p_1$ into $-p_4$. One is therefore left with six diagrams whose sum can be shown, by the same techniques as at one loop, to be equal to the two loop contribution to the eikonal amplitude. Note that, as expected and as at one loop, the leading diagrams are the ladder and crossed ladder diagrams.

\subsec{Graviton dominance}

In particular, in the context of SUGRA, we  find {\it graviton dominance} in this high-energy, long distance region.  The essential reason for this is that the graviton couples to the stress tensor, giving a coupling that grows with energy faster than couplings of other states.  For long-range behavior, one of course only considers massless exchanged states.

To outline this argument further, first consider replacing an exchanged graviton in \gravladd\ by a scalar with a non-derivative coupling.  In this case, the leading energy dependence of the vertex changes as $E^2\rightarrow M_D^2$, times a dimensionless coupling.  Only for a scalar with a coupling to a two-derivative quantity, as with the stress tensor, would the scalars compete in the high-energy regime we consider.  
We could likewise consider exchange of a vector particle, coupled to a current; here the parametric replacement is $E^2\rightarrow E M_D$, up to dimensionless factors.  

A similar story holds for exchange of fermions, such as gauginos or gravitinos.  In these cases, the propagator of the exchanged particle behaves as ${\displaystyle \not k}/k^2$.  Thus, since there is an extra power of the exchanged momentum in the diagram, there is one fewer power of the energy $E\gg k$, resulting from change of the vertex factor.  (Of course, in this case the other particle exiting the vertex must also be fermionic if the incoming particle was bosonic.)

Also higher p-form fields couple with lower powers of energy at the vertex than gravitons. The reason is that due to the antisymmetry in the propagators and/or vertices, the external momenta are contracted in such a way that only one power of the high momentum appears.

This explains the quantitative sense in which gravity dominates over other SUGRA states, far away from the BPS limit.

\newsec{Physical consequences}

We have argued for dominance of  long distance physics, near a saddlepoint, in the high energy limit.  This occurs in the eikonal regime, where $\chi\roughly>1$, with the eikonal phase $\chi$ given in \eikphase.  As we have noted, once we enter that regime, given in terms of impact parameter as
\eqn\eikimp{b\roughly< (E/M_D)^{2/(D-4)}/M_D\ ,}
which also corresponds to $q\roughly> 1/b$, the supergravity loop expansion about flat space becomes problematic -- it is analogous to the expansion of the exponential in our toy integral, \toyint.  

\subsec{Semiclassical amplitudes}

The eikonal approximation is of course closely connected with the semiclassical approximation.  Indeed, one useful analogy for thinking about high-energy, long-range gravitational scattering is as similar to Earth-Moon scattering (for the moment ignoring bound state effects).  In the Earth-Moon system, the two interacting bodies have ultraplanckian energies.  However, the momentum transfers of individual gravitons is very small.  It is through exchange of many such gravitons that the Earth has an appreciable effect on the moon, which is well-described through the classical field.

In the context where the energy is large due to ultrarelativistic incoming particles, and in the large distance limit, a similar story seems to be play out.  That raises the question of the classical description.  It is of course known that the classical geometry of an ultrarelativistic particle is an Aichelberg-Sexl\refs{\AiSe} shockwave.  Moreover, there is an exact solution describing two such colliding shockwaves, in the region outside of the future lightcone of the intersection of the two shocks.  This follows from the form of the Aichelburg-Sexl solution, 
\eqn\as{ ds^{2} = -dx^+ dx^- + dx^{i} dx^{i} +\Phi(\rho) \delta (x^-) dx^{-2}}
with
\eqn\fd{\eqalign{ \Phi(\rho) &=  -8G_D\mu\ln\rho\quad ,\quad D=4\ ,\cr
&= {16\pi G_D\mu\over \Omega_{D-3} (D-4) \rho^{D-4}}\quad ,\quad D>4\ ,}}
$x^\mu=(x^+,x^-,x^i)$, $\rho^2=x^ix^i$, and $\mu=E/2$.  This is flat in the regions both in advance of and behind the shocks.  This means that solutions describing incoming shocks from opposite directions can be trivially glued together in their common advance region, prior to the collision of the shocks, to give an exact classical solution.

We expect the full classical solution (perhaps with some smearing\refs{\GiRy}) to correspond to the saddlepoint about which we expand to compute quantum amplitudes.  Namely, once we reach the region \eikimp, expanding about flat space is no longer a particularly useful picture.  Instead, one should find the classical geometry of the colliding shock waves, and then perform an expansion in perturbations about that.  This corresponds to a resummation of the loop amplitudes to give the classical geometry.

The full classical geometry is not known, even for large impact parameters, in the future of the collision.  However, for large impact parameters, we do expect a test-particle approximation to furnish a first approximation.  Namely, if a test particle scatters in the Aichelburg-Sexl solution, and we neglect its backreaction on that geometry, then it is simply deflected by a classical deflection angle, given parametrically in terms of the energy and impact parameter by \classang.  So, we might treat as an approximate solution the collision of two such shocks, where we assume the center of each shock is deflected along a corresponding geodesic in the gravitational field of the other shock.  The small expansion parameter in this context should then be the deflection angle, $\theta_c$.  Note from \classang\ that this is small for $b\gg R(E)$.  

Indeed, such an expansion scheme seems essentially similar to that of the worldline effective theory approach of \refs{\GoRo}, used to derive solutions for neutron star or black hole collisions.  We suggest that a systematic such parallel can be derived, and this is currently under exploration\refs{\GSP}. 

For impact parameters $b\roughly< R(E)$, with $R(E)$ the Schwarzschild radius given in \schwr, it is known that such a collision produces a trapped surface, hence a black hole\refs{\Penrose,\EaGi} -- that is, gravity becomes strongly coupled.  We can also see this is the region where subleading corrections to the eikonal amplitudes become important, since the eikonal amplitudes arose to leading order in an expansion in $-t/s$, and by \classang,
\eqn\expparam{{-t\over s}\sim \left[{R(E)\over b}\right]^{2(D-3)}\ .}

It would be useful to more systematically understand quantum corrections, and where one expects a valid perturbative expansion.\foot{One useful way to systematize this is based on expanding about the saddlepoint in the functional integral over metrics and other fields.}  First, note that from the perspective of the expansion about flat space, in the region \eikimp\ one can think of the eikonal sum as dominated by the terms of order $N\sim \chi$ in the expansion of the exponential\refs{\GiPo}, and with the momentum transfer $q$ roughly divided between the lines, this corresponds to a characteristic momentum 
\eqn\linemom{k\sim {q\over N} \sim {1\over b} }
in each of the exchanged graviton lines.  That is, the total momentum transfer $q$ undergoes fractionation into a large number of softer momentum transfers.  

Another interesting question is whether nonrenormalizable interactions, {\it e.g.} of the form $\phi\partial^n\phi$, make important contributions in any regime.  We suggest that such pointlike (thus short distance) interactions might be avoided in the high-energy regime by controlling the impact parameter, so that for example the wavefunctions of the incident states have very small overlap.  However, we leave this question for future investigation.

\subsec{Consequences of momentum fractionation}

This picture (and  improvement of it based on the resummation indicated above) apparently has several interesting consequences.

\subsubsec{Tidal excitation}

The relative softness of the momentum transfer figures into the role of excitation of extended or composite objects, such as strings, in gravitational scattering.  Specifically, if the scattered objects are strings,\foot{Similar statements are expected for any other extended object, {\it e.g.} hydrogen atoms.} there can be a large net momentum transfer without exciting them.  However, below a certain threshold impact parameter, the tidal force that a string experiences when crossing the gravitational shock wave of the other string will excite it.  This basic effect was originally seen as diffractive excitation in \refs{\ACVone} before being given the tidal interpretation in \refs{\LQGST}.  There has been some confusion about this effect, and it has even been suggested that it might be responsible for information escaping a black hole\refs{\Venez}, but \refs{\LQGST,\GGM} argued that this kind of excitation takes place on time scales that are parametrically longer than horizon formation.

\subsubsec{Soft radiation}

Directly related to this, since at any one graviton vertex the momentum transfer is small, $\sim 1/b$, we expect this to set the transverse momentum scale of soft radiation.  
Properties of this radiation can be estimated by methods of \refs{\Weinsoft}; for example \refs{\GiSr} gave an estimate of the resulting absorption from the elastic channel due to such emission.

\subsubsec{Factorization scale}

Another consequence of this is that if one considers black hole production in hadronic collisions, which could be experimentally relevant in TeV-scale gravity models\refs{\GiTh,\DiLa},\foot{For a review with some further references, see \refs{\SGPascos}.}
 the relevant factorization scale in the context of QCD processes is $q\sim 1/b$ -- this was noted and used in \GiTh\ for computing black hole production cross sections, and has been further elaborated by \refs{\Empar,\Rych}.  
 
\subsubsec{Asymptotic safety}

We suggest that the softness of individual momentum transfers also has an important consequence for the asymptotic safety program\refs{\WeinAS}.\foot{For a review and further references, see \refs{\ASrev}.}  The basic tenet of that program is the existence of a fixed point in the high-energy limit.  

If we ask about the possible physical meaning of such a fixed point, a proposed answer appears in \WeinAS:  one should define the couplings in terms  of physical reaction rates, at a given energy scale.  Correspondingly, we should be able to probe such a fixed point through {\it physical} processes, such as high energy scattering.  A familiar example is the asymptotically safe and free coupling of QCD, which can be probed via, {\it e.g.}, deep inelastic scattering. 
 Indeed, one might expect that such a phenomenon is only physically meaningful and relevant to the extent it can be probed through such  processes.  
 
 For such a sharp physical test, the obvious regime to investigate is high-energy scattering, with large momentum transfer.  
 However, we have  argued that the regime of high energies and large momentum transfer is dominated by processes very far from single graviton exchange!  In the physical processes we have described, one is apparently {\it not} sensitive to the single graviton vertex and thus to the graviton coupling at high momentum transfer, since a large total momentum transfer arises from the exchange of multiple soft gravitons, each of whose vertices depends on the gravitational coupling in the low momentum transfer regime, $k\sim 1/b$.  Correspondingly, in quantum gravity there is an apparent obstacle to physically probing distances shorter than the Planck scale via scattering.

Thus, we see no clear way to relate the gravitational coupling at a cutoff scale $\Lambda\rightarrow\infty$ to physical high-energy scattering.  Instead, we enter a regime where, to the extent that field theory is a good description, the physics is dominated by multi-graviton exchange.  
It should be noted that the dominant scattering process in this high-energy limit has a many-particle final state; that is, $2\rightarrow2$ scattering is highly suppressed, particularly when one enters the strong-gravity regime, $b\roughly<R(E)$ (see \refs{\GiSr,\GiPo} for further development of this point).  We might conjecture that the vanishing of $G_D$ seen in the formal calculations with $\Lambda\rightarrow\infty$ investigated in the literature is connected to the relative unimportance of single graviton exchange in the high-energy limit; this is a question worth further exploration.

\subsubsec{Regge behavior}

\Ifig{\Fig\hdiag}{The H-diagram.}{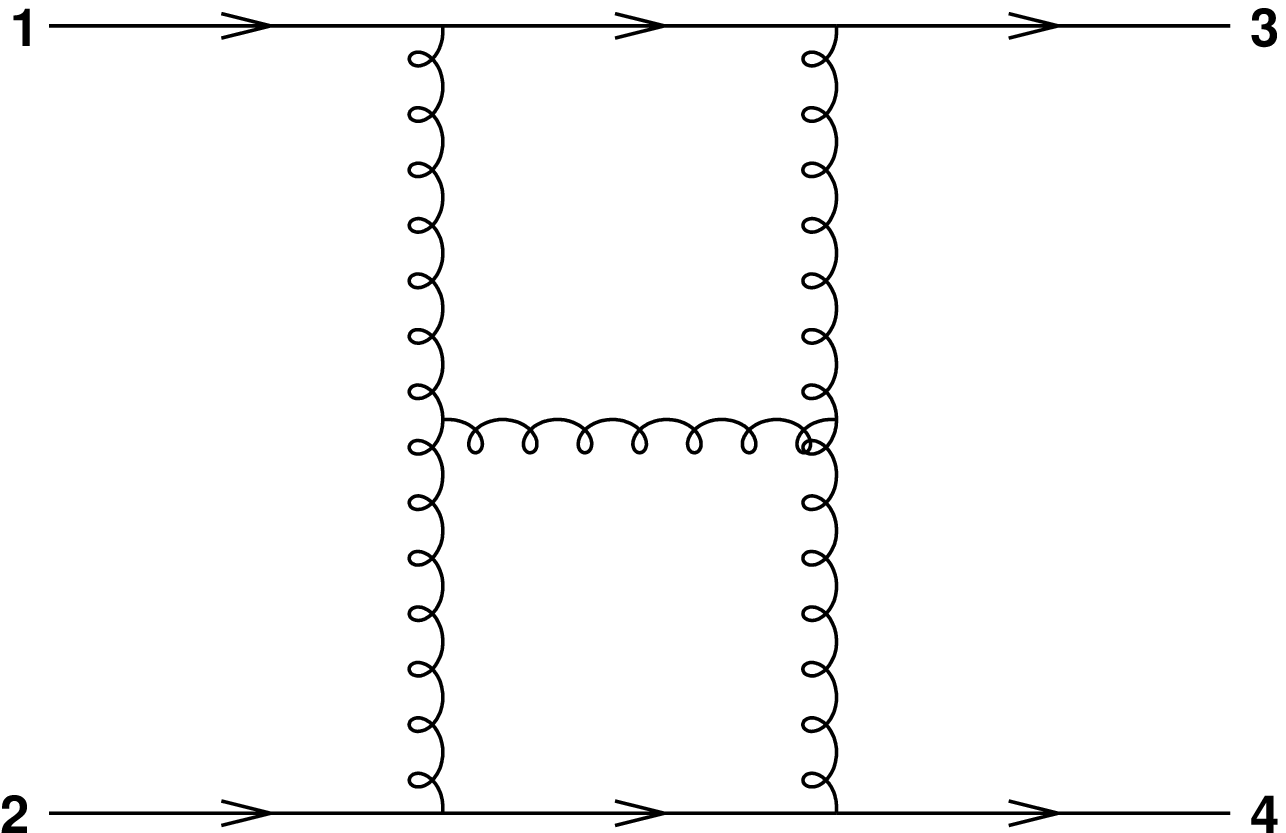}{3}

\Ifig{\Fig\rdiag}{A t-channel ladder diagram that would contribute to Regge behavior.}{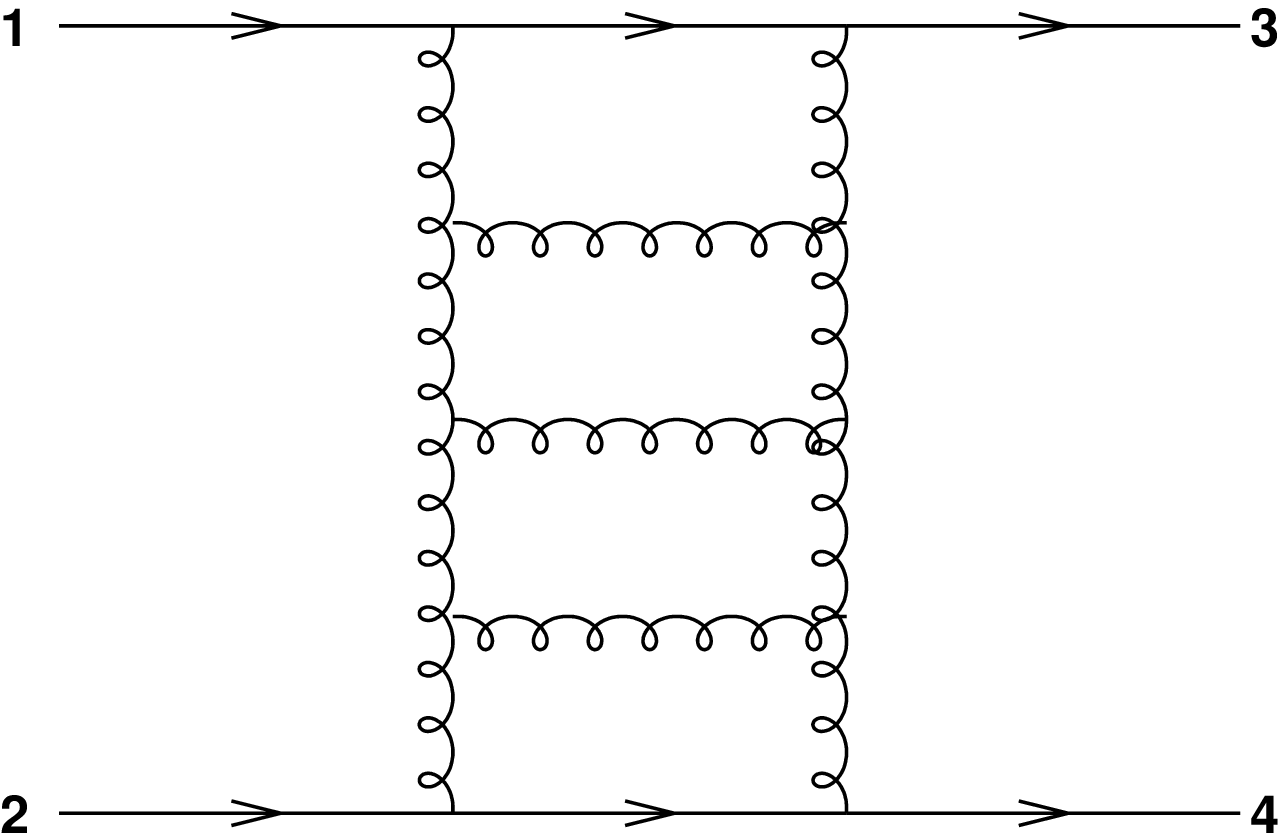}{3}

Another related matter is the possible reggeization of gravity\refs{\Schnone,\Schntwo}, which has been considered an open question.  To address this, consider subleading diagrams to the eikonal diagrams.  One important class of such diagrams are the H-diagrams, pictured in \hdiag.  In particular, these were found in section five to be subdominant terms in the $-t/s$ expansion, in agreement with general arguments presented in \refs{\ACVthree,\ACVfour,\GiPo}.   Regge behavior could make important contributions to high-energy scattering if ladders such as shown in \rdiag\ become important.  But, such diagrams will only become relevant when the momentum flowing through the  ladder is sufficiently large, with the argument being similar to that for the role of ladders in the eikonal approximation.  But, we have argued that the dominant contribution to high-energy scattering in the regime \eikimp\ instead involves many soft momentum transfers of size \linemom\ through the $N\sim \chi$ lines.  If these are the dominant diagrams, then the ladders of \rdiag\ would appear to be relatively highly suppressed, and the high-energy scattering does not probe possible Regge behavior.  

\subsubsec{Strong gravity/perturbative breakdown}

From the point of view of perturbation theory about flat space, the $H$ diagrams themselves only become important in the regime $-t\sim s$, or $b\sim R(E)$.  This is, of course, the regime where classically a black hole would form.  Indeed, consider the class of diagrams consisting of graviton tree diagrams, with all external gravitons attached to one or the other of the two high-energy colliding particles.  In this way we can think of the colliding particles as sources.  A similar class of diagrams was considered by Duff\refs{\Duff}, with an external source which was a single massive particle.  In that case, Duff showed that the sum over graviton trees build up the Schwarzschild metric.  This indicates that the same should be true here, namely that such tree diagrams should be building up the classical geometry of the colliding Aichelburg-Sexl shocks.  When the impact parameter reaches $b\sim R(E)$, and a black hole forms, this series diverges, like the Schwarzschild metric, thought of as a perturbation about flat space, diverges at the horizon.  

Of course, by the time this divergence occurs, one has already passed beyond the region where one should resum, and perturb about the  colliding Aichelburg-Sexl geometries, corresponding to the necessity of performing the eikonal sum.  So, a better framework for describing this regime is precisely the perturbation theory in $-t/s$ about this lowest-order metric, a la \GoRo, suggested above.  However, we expect that the basic feature of divergence of this perturbation series remains, in the region $-t\sim s$, or $b\sim R(E)$, where a black hole forms.  The black hole solution cannot be thought of as a small perturbation of colliding Aichelburg-Sexl shock waves.

In short, at first one expands about flat space; in the eikonal region, one must resum and expand around a classical geometry corresponding to that of the collision of two Aichelburg-Sexl geometries, which mutually deflect; but by the time the classical black hole region is reached, the classical solutions develop horizons (and singularities), and are not perturbatively related to either of the preceding geometries.  

\subsec{Nonrenormalizability, unitarity, and nonperturbative mechanics}

This discussion appears to carry some important lessons for quantum gravity, elaborating on points outlined in \refs{\BeyP}. 

First, in studying the high-energy dynamics, it appears that the central issue is {\it not} necessarily renormalizability.  While a naive approach to the loop expansion in the high-energy regime, like the expansion \powerser\ of the toy integral \toyint, involves arbitrarily divergent behavior, which must be regulated, what we have argued is the more correct dynamics does not centrally involve such short-distance physics.  It is true that a perturbative expansion about flat space, or about classical geometries of colliding Aichelburg-Sexl metrics, can have UV-divergent corrections.  But, in the regime of ultrahigh energy scattering, one is apparently only probing dynamics where individual gravitons carry momenta well below $M_D$, and correspondingly one is not probing short distances.   This physics appears largely insensitive to the details of a regulator $\Lambda\sim M_D$.   This is the gravitational version of a UV-IR correspondence.  These statements pertain both to the eikonal regime, and to the black hole regime; their black hole version has been previously emphasized for example in \refs{\BaFi\GiTh\EaGi\Banks-\LQGST,\GGM}.

This discussion neglects aspects of the black hole regime, where one might think that in some sense one is probing physics at distances $\sim 1/M_D$, when particles ``reach the singularity,"  or at the end stage of black hole evaporation.  However, the black hole regime presents what has been argued\refs{\BeyP} to be a much more serious and central problem -- a loss of unitarity.

Specifically, if one finds that resummations reproduce the classical geometry, and if there is such a resummation possible for $b\roughly<R(E)$, the classical geometry should be that of a black hole.  Then, the quantum amplitudes might be expected to be given by that saddlepoint, and perturbations about it.  However, we know that a naive analysis of this problem violates unitarity badly, as first showed by Hawking in arguing for loss of quantum information\refs{\Hawkunc}.  Indeed, it has also been observed that there is no way for gravitational scattering to be unitary in this regime, {\it i.e.} for information to be returned, in a theory described as local quantum field theory perturbations about a background.\foot{For reviews, see \refs{\Astrorev,\SGinfo}.}  

This argument strongly suggests that in order for the theory to be unitarized in the regime corresponding to classical black hole formation, the dynamics needs to have some intrinsic nonlocality, at least as described with respect to the semiclassical geometry.  This nonlocality apparently must operate over distances of order $R(E)$, to permit information escape from the horizon of  a macroscopic black hole, by the necessary time scale $t\sim R(E) S(E)$ originally found by Page\refs{\Page}, where $S(E)$ is the black hole entropy.

By this logic, the problem of unitarity appears to be  a deeper issue of quantum gravity than that of nonrenormalizability.  If we accept that the as yet unknown mechanisms of nonperturbative quantum gravity {\it must} achieve such unitarization, this seems like a very important guide to understanding those mechanisms\refs{\LQGST,\GiddingsSJ,\GiddingsBE}.\foot{A first approach to understanding such mechanisms is to seek a parametrization of the ``correspondence boundary" where they give important corrections to local field theory; this motivates formulation of the locality bounds of \refs{\GiddingsPT,\GiddingsSJ,\GiddingsBE}.} We stress that this is apparently {\it not} a short distance problem; if the lessons of the discussion of the information problem are to be taken seriously, modifications of physics at short distances $\sim 1/M_D$ cannot achieve such unitarization.  This guide thus seems particularly important in suggesting that the requisite mechanics represents a significant departure from local quantum field theory, at large distances, at least in the unfamiliar realms of black holes.

\bigskip\bigskip\centerline{{\bf Acknowledgments}}\nobreak
We thank A. Denner, G. Giudice, L. Magnea, J. Polchinski, R. Porto, S. Pozzorini, D. Trancanelli, G. Veneziano, J. Wells, and E. Witten for valuable discussions, and C. Gomez for discussions on reggeization.   We would particularly like to thank L. Dixon for supplying information on gravitational loop amplitudes.  SBG gratefully acknowledges the kind hospitality of the CERN theory group, where this work was initiated. M.S-S. thanks the KITP, Santa Barbara, for hospitality during part of this work;
his visit and research was supported in part by the National Science Foundation under Grant No. NSF PHY05-51164.
The work of SBG was supported in part by the Department of Energy under Contract DE-FG02-91ER40618.

\listrefs
\end